# Coherent Spin Manipulations of a Polarized Beam
# With a Localized RF Magnetic Field

**Dennis Sivers**

*Spin Physics Center*
*University of Michigan*
*Ann Arbor, MI*

*Portland Physics Institute*
*4730 SW Macadam, suite 101*
*Portland, OR 97239*

**Abstract**

The coherent manipulation of spin observables in storage rings provides opportunities to test the application of some fundamental dynamical principles. In this context, it is possible to confirm, using gauge invariance and Lorentz invariance, a conjecture framed by A.M. Kondratenko concerning the "natural" or "intrinsic" resonance strength applicable to a spin rotation from a controlled Froissart-Stora sweep with an RF dipole magnet. The discussion includes a brief treatment of the "forced" component of the resonance strength associated with the effect of the betatron oscillations induced by the operation of the RF dipole and a discussion of the effective resonance strength as a function of betatron tune.

Draft modified 8/15/07

**I. Introduction**

The idea that fundamental particles can have spin is associated with the space-time symmetry of Poincare. [1] The three-vector identifying the spin orientation of a moving particle with finite mass and a 4-momentum $P^\nu$ can be defined by transforming the Wigner-Pauli-Lubanski tensor [2]

$$W^\nu = \frac{1}{2} g^{\nu\lambda} \varepsilon_{\lambda\rho\sigma\mu} J^{\rho\sigma} P^\mu \qquad (1.1)$$

to the particle rest frame by means of a rotationless Lorentz-boost. This implies that, under an arbitrary Lorentz transformation, $\Gamma^\mu{}_\nu$, that transforms the 4-momentum of the particle

$$P'^\mu = \Gamma^\mu{}_\nu P^\nu \qquad (1.2)$$

the transformation of the spin 3-vector requires the product of three Lorentz transformations known as the Wigner rotation. [3]

Since the particle spin degrees of freedom are necessarily quantized in order to be consistent with CPT-invariance of the underlying dynamics, [4] the complicated Lorentz transformation properties of the spin 3-vector require that strict attention be paid to the conventions regarding the basis in which the quantization is specified. The advantages of specifying spin degrees of freedom in the helicity basis are summarized in the comprehensive work of Jacob and Wick. [5] For our limited purposes here, we note that helicity states for individual particles are separately boost covariant and rotationally covariant so that individual particle quantum states can be classified by the appropriate representation of the Lorentz group. In the helicity basis, particle states for a particle of mass m, 3-momentum $\vec{p}$, and spin $\vec{s}$, $|m; \vec{p}, \vec{s}\rangle$ are quantized into 2s+1 helicity states

$$|m; \vec{p}, \vec{s}\rangle = \sum_\lambda a_\lambda |m; \vec{p}, \lambda\rangle \qquad (1.3)$$

where the helicity, $\lambda$, is a pseudoscalar observable, $\lambda = -s, -s+1, ...,$ that is odd under parity. Under a general Euler-angle rotation, the momentum vector and the spin vector of the particle both rotate together and the rotated spin projections on the helicity axis are given by the operator

$$D^J_{\lambda\lambda'}(\alpha, \beta, \gamma) = e^{i\alpha\lambda} d^J_{\lambda\lambda'}(\beta) e^{i\gamma\lambda'} \qquad (1.4)$$

with J=s. Following the Jacob-Wick conventions preserves the full rotational and Lorentz invariance of the theory. These conventions involve defining the separation of the longitudinal and transverse degrees of freedom for the Wigner-Pauli-Lubanski tensor for each particle in a definite helicity state. For the particle 4-momentum

$$P_i^\nu = m(\cosh Y_i, 0, 0, \sinh Y_i) \tag{1.5}$$

where $Y_i$ defines the particle's rapidity and we have arbitrarily specified that the momentum is along the $\hat{z}$ axis. This specification gives the separation

$$W_{iL}^\nu = ma_\lambda \left(\frac{\lambda}{\cosh Y_i}, 0, 0, \frac{\lambda}{\sinh Y_i}\right)$$
$$W_{iT}^\nu = m[s(s+1) - \lambda^2 |a_\lambda(Y_i)|^2]^{\frac{1}{2}} (0, \cos\phi_i, \sin\phi_i, 0) \tag{1.6}$$
$$|a_\lambda(Y)_i|^2 = |a_\lambda|^2 / \cosh^2 Y_i \sinh^2 Y_i$$

In the helicity basis, information concerning the transverse degrees of freedom of spin is thus carried by complex phases in accordance with the algebra of the rotation group, SO(3), or its covering group SU(2).

This separation is required for identifying the appropriate degrees of freedom for specifying the couplings of the vector bosons of the standard model (photons, gluons, W and Z bosons) to the fundamental fermions (neutrinos, charged leptons, quarks) in conformance with the gauge symmetries of the theory. [6] In particular, the gauge invariance of electrodynamics for the coupling of a charged particle with spin to electromagnetic fields requires that we separate the magnetic moment of the particle into two components

$$\mu = \frac{q}{m}(1+G) \tag{1.7}$$

where $q/m$ is the Dirac moment and $Gq/m$ is the particle's anomalous magnetic moment. For a fundamental fermion, the magnetic anomaly G can be calculated to a high degree of accuracy using the virtual fluctuations of the quantum field theory. Therefore, the fact that the electron's magnetic anomaly

$$G_e = 1.15967 \times 10^{-3} \tag{1.8}$$

can be accurately calculated in the standard model provides strong support both for gauge invariance and for confirming that the electron is a pointlike particle. In contrast, the large anomalous magnetic anomaly for the proton

$$G_p = 1.792847 \tag{1.9}$$

shows that it has considerable internal structure. [7]

Having specified a consistent basis for particle spin observables, we are interested in the coherent spin manipulation of polarized beams with radio-frequency (RF) magnetic fields. A polarized beam can be defined as an ensemble of known particles such that the beam density matrix has a nonzero expectation value for the particle spin operator, $\Sigma$.

$$\vec{S} = \langle \vec{\Sigma} \rangle = tr[\rho \Sigma] \tag{1.10}$$

It is assumed that the particle's spin and anomalous magnetic moment are known. In what follows, we will primarily be discussing protons and electrons, spin-$\frac{1}{2}$ fermions with the anomalous magnetic moments given above and deuterons, a spin-1 bound state of a proton and neutron with magnetic anomaly given by

$$G_d = -0.14297 \tag{1.11}$$

However, we will keep the approach general. It is important to keep in mind that the three particles named above have a wide range of values for the anomalous moments. The coherent spin manipulation of a polarized beam can provide a powerful diagnostic tool in an accelerator or storage ring. This can be seen by expressing the relativistic form of the spin vector, $\vec{S}$, in equation (1.10) in terms of the operators $P^\upsilon$ and $W^\upsilon$. For a pure quantum state of the beam that is a coherent superposition of particle helicity states, this would give the expectation values

$$\langle P^\upsilon \rangle = m(\cosh\langle Y \rangle, 0, 0, \sinh\langle Y \rangle)$$
$$\langle W^\nu \rangle = m[s(s+1)]^{\frac{1}{2}}(\cos\langle\theta\rangle\sinh\langle Y\rangle, \sin\langle\theta\rangle\cos\langle\phi\rangle, \sin\langle\theta\rangle\sin\langle\phi\rangle, \cos\langle\theta\rangle\cosh\langle Y\rangle) \tag{1.12}$$

where the angle $\langle \theta \rangle$ is defined by

$$\cos\langle\theta\rangle = \frac{\langle\lambda\rangle}{\cosh\langle Y\rangle\sinh\langle Y\rangle}$$
$$\sin\langle\theta\rangle = \pm(1-\cos^2\langle\theta\rangle)^{\frac{1}{2}} \tag{1.13}$$

and we can define the orientation of the mean spin 3-vector

$$\hat{S} = \frac{\vec{S}}{[s(s+1)]^{\frac{1}{2}}} = (\sin\langle\theta\rangle\cos\langle\phi\rangle, \sin\langle\theta\rangle\sin\langle\phi\rangle, \cos\langle\theta\rangle) \tag{1.14}$$

relative to the mean particle momentum by taking the limit $\sinh\langle Y\rangle \to 0$ for both the momentum and the Wigner-Pauli-Lubanski tensor in (1.12) using measurable experimental quantities that can be determined with appropriate polarimetry. Typical beam conditions will fall short of a pure quantum state but local rotations preserve the purity of quantum states and studying the coherent effect of repeated localized rotations and measuring spin observables therefore provides a powerful protocol for studying beam dynamics. It is convenient for further calculations involving beam parameters to retain the separation between the transverse and longitudinal components of the Wigner-Pauli, Lubanski tensor. The transverse components of this tensor

$$\langle W^\mu \rangle_T = m\sqrt{s(s+1)}(0, \sin\langle\theta\rangle\cos\langle\phi\rangle, \sin\langle\theta\rangle\sin\langle\phi\rangle, 0)$$
$$\langle W^\mu \rangle_T = m(0, \vec{S}_T) \tag{1.15}$$

transform under Lorentz boosts <u>only</u> by means of the variation of $\sin\langle\theta\rangle$ given in (1.13). Assuming that $\langle\lambda\rangle$ is fixed because of parity invariance, the longitudinal components can be written

$$\langle W^\mu \rangle_L = m\langle\lambda\rangle\sqrt{s(s+1)}(\frac{1}{\cosh Y_i}, 0, 0, \frac{1}{\sinh Y_i}) \tag{1.16}$$

In the large Y limit this gives

$$\lim_{Y_i \to \infty} \langle W^\mu \rangle_L = m\langle\lambda\rangle\sqrt{s(s+1)}e^{-|Y_i|}(1, 0, 0, \pm 1) \tag{1.17}$$

where the sign of the z-component depends on the sign of the rapidity $Y_i$. The nonrelativistic limit can be specified in the form

$$\lim_{Y_i \to 0} \langle W^\mu \rangle_L = m\sqrt{s(s+1)}(\langle\lambda\rangle, 0, 0, \cos\langle\theta\rangle)$$
$$\cos\langle\theta\rangle = \lim_{Y_i \to 0} \frac{\langle\lambda\rangle}{Y_i} \tag{1.18}$$

It is important to keep in mind here that $\langle W^\mu \rangle_L$ cannot be measured with conventional polarimetry but instead requires measurement of two-spin observables in order to be determined experimentally. This occurs because the combination of finite symmetries and rotational invariance provides significant constraints on what spin observables can be measured. Specifically, in processes that preserve parity, only polarizations transverse to the scattering plane can be measured in the scattering of a polarized beam with an unpolarized target. The measurement of helicities thus requires both polarized beam and polarized target in a scattering process. Therefore, to describe beam spin observables in a storage ring without Siberian snakes where the orbit of the beam is located in the x-z plane in terms of the angles specified above for the Wigner-Pauli-Lubanski tensor, requires a polarization measurement, with

$$P = \hat{e}_y \cdot \hat{S} = \sin\langle\theta\rangle\sin\langle\phi\rangle \tag{1.19}$$

For the purpose of discussing such measurements for a polarized beam, it is convenient to use a basis for the operator $\Sigma$, in Eq. (1.10), so that the eigenvalues for $\Sigma_y$ are diagonal. A convenient choice involves the cylindrical coordinates:

$$\Sigma_1 = \vec{\Sigma} \cdot \hat{r}, \Sigma_2 = \vec{\Sigma} \cdot \hat{p}, \Sigma_3 = \vec{\Sigma} \cdot \hat{e}_y \qquad (1.20)$$

where we have specified the particle beam orbit in the ring to be in the x-z plane. For a magnet located at a specific location in the ring, the other two directions are then determined. For an eigenstate of $\Sigma_3$ the density matrix is then diagonal and we will define the vector polarization in this basis by

$$P = n_s - n_{-s} \qquad (1.21)$$

in terms of the number densities of the maximal states. For spin-$\frac{1}{2}$ particles, this is the only polarization observable possible. For a spin-1 particle, such as the deuteron, we can also define the so-called "tensor" polarization as

$$P_T = 3n_0 - 1 \qquad (1.22)$$

In what follows, we will describe spin manipulations in terms of polarizations. If no additional label is included, we are referring to the vector polarization.

A significant aspect of the application of this program involves the concept of a Froissart-Stora sweep.[8] The acceleration of a polarized beam through a spin resonance of known strength, $\varepsilon$, in an accelerator was shown in Ref. 8 to give a relationship between the initial polarization, $P_i$, and the final polarization, $P_f$, of the form

$$P_f = P_i \{2\exp(-\pi |\varepsilon|^2 / 2\alpha) - 1\} \qquad (1.23)$$

where $\alpha$ is the acceleration ramping speed. For a weak resonance and/or rapid crossing speed, $\pi |\varepsilon|^2 / 2\alpha \ll 1$, the polarization remains essentially unchanged. On the other hand, for a strong resonance and a sufficiently slow ramp speed, $\pi |\varepsilon|^2 / 2\alpha \gg 1$, the polarization will coherently rotate through an angle $\pi$ following an adiabatic path of pseudostable spin orientations so that $P_f / P_i \cong -1$.

The concept of accelerating a particle beam through a resonance can be extended to allow coherent spin manipulation of beams of fixed energy [9] by using a frequency-controlled RF magnet to artificially induce a spin resonance, with resonance condition,

$$f_r = f_c (k \pm \nu_s) \qquad (1.24)$$

by giving the particle a kick with a magnetic field on each turn around the ring. In (1.24), $f_c$ is the particle's circulation frequency, k is an integer and $\nu_s$ is the spin tune. Ramping the magnet through a frequency range, $\Delta f$, that includes a single resonance over a time $\Delta t$ gives the Froissart Stora formula in the form

$$P_f = P_i \{2\exp[\frac{-(\pi\varepsilon f_c)^2}{\Delta f / \Delta t}] - 1\} \tag{1.25}$$

These sweeps can be done either with a solenoid magnet, in which case the rotation induced is around the beam axis, or with a dipole with a rotation axis in the plane of the beam. The control of both the strength of the magnetic field and the sweep parameters allows coherent rotations of polarization with mean angles between 0 and $\pi$. In the case of a solenoid magnet, the rotation of the Polarization observables defined in (1.23) – (1.25) above is around the $\hat{z}$ axis and is consistent with the separation of longitudinal and transverse spin degrees of freedom as given in (1.15)-(1.18). For a dipole magnet, however, care must be taken in preserving this separation, because the magnetic field deflects both spin and momentum. Studies involving Froissart-Stora sweeps with local magnetic fields have been done with horizontally and vertically polarized protons at the IUCF Cooler ring [9-16] and COSY [17,20,21] deuterons with both vector and tensor polarizations at the IUCG Cooler ring [18,19] and at COSY [17,20,21] and with horizontally polarized electrons at the MIT-Bates storage ring [22]. An RF dipole with controlled frequency has also been used at the AGS to enhance the strength of an intrinsic resonance so that the two coupled resonances could be crossed adiabatically. [23]

The use of RF magnets to coherently rotate the spin of polarized beams of fixed energy has opened up potential new avenues for manipulating beam properties. Chao [24] has recently developed a matrix formalism that extends the Froissart-Stora formula to deal with partial resonance crossings or with resonance crossings involving nonlinear ramp profiles. In addition, Kondratenko [25] has suggested resonance-crossing profiles for crossing weak resonances that may improve on existing techniques for beam acceleration. The detailed adjustments necessary for these improved techniques can be tested on beams of fixed energies in storage rings using the spin resonances induced by localized magnetic fields. Control of both frequency and strength of the RF magnet along with control of the acceleration ramp allows for many variations [26] of "rotator-aided" resonance crossing for intrinsic and imperfection resonances.

An important component of the application of these precision techniques for spin manipulation is the firm understanding of the determination of the resonance strength for the induced spin resonances generated by the operation of the RF magnet. Conventional calculations [27] give the results for the resonance strength induced by a localized solenoidal magnet (S) or for a localized dipole magnet (D) operating in a ring without Siberian Snakes as

$$\varepsilon^S = \frac{q(1+G)}{2\pi p\sqrt{2}} \int B_T^{RMS} dl \tag{1.26}$$

and

$$\varepsilon^D = \frac{q(1+G\gamma)}{2\pi p\sqrt{2}} \int B_T^{RMS} dl \tag{1.27}$$

In these equations and in the remainder of the section, observe that the Lorentz boost from the rest frame of the particles in the beam is indicated by $\gamma = \cosh Y$ in the expressions for resonance strength. We will use both types expressions to describe the magnitude of Lorentz boosts as appropriate from this point.

A.M. Kondratenko, [28] while agreeing with (1.26) for the strength of the resonance generated by a localized RF solenoid, has presented compelling general arguments that indicate

$$\lim_{G \to 0} \varepsilon^D = \mathrm{O}(G\gamma) \tag{1.28}$$

for the resonance strength generated by a localized RF dipole magnet. These arguments directly challenge the result (1.27) and this challenge represents the Kondratenko conjecture mentioned in the abstract. In this paper, we confirm Kondratenko's conjecture and present a calculation of the "natural" or "intrinsic" resonance strength of a dipole magnet to demonstrate

$$\varepsilon^D = \varepsilon_K^D = \frac{qG\gamma}{2\pi p\sqrt{2}} \int B_T^{RMS} dl \tag{1.29}$$

This expression is valid when the localized RF dipole operates in a ring without Siberian snakes and the magnitude of the magnetic field is limited by some stability criteria. As has been indicated above, repeated coherent rotations with a localized RF dipole present some challenges in maintaining the separation of longitudinal and transverse degrees of freedom for spin observables since the magnetic field rotates both the momenta and the spins of the beam particles. To validate the result (1.29) we demonstrate that our calculation respects gauge invariance, Lorentz invariance, and invariance under finite symmetries. The difference between (1.29) and (1.27) is not large when $G\gamma \gg 1$ but can be important for specific precision applications. The known values for the anomalous moments (1.8), (1.9) and (1.11) demonstrate that this difference is important for a much larger range of energies for beams of electrons and deuterons than for proton beams. We interpret the value of the Kondratenko resonance strength given in (1.29) as an "intrinsic" strength as defined by Lee in Ref. [29] in that it is based only on the spin rotation angle generated by the local fields of the RF dipole magnet. In a specific experimental application, we must consider, in addition, the forced betatron and synchrotron motion produced by the momentum deflection generated by operation of the dipole. These oscillations necessarily induce an additional component to the resonance strength associated with the RF dipole, called the "forced" resonance strength by Lee, since it is based on the spin rotation that occurs in the remainder of the ring. This component can be either positive or negative. When appropriate, to avoid confusion with other applications of the word "intrinsic", we will also refer to Lee's "intrinsic" resonance strength as the natural resonance strength associated with the local fields.

The remainder of the paper is organized as follows: Section II presents the calculation of the intrinsic resonance strength appropriate for a controlled Froissart-Stora sweep with a RF dipole magnet. We show there that the result, (1.29) above, is required by gauge invariance, boost invariance, rotational invariance and parity invariance of local electromagnetic interactions. Section III deals with the issue of including the "forced" component of spin rotation. For a Froissart-Stora sweep that crosses only the induced resonance, this topic can be approached through resonance mixing. Section IV concludes with a brief outline of the resultant variation as a function of betatron tune for the effective resonance strengths observed using controlled Froissart-Stora sweeps involving RF dipole magnets.

## II. Resonance Strength for a Local RF Magnetic Field

The starting point for any discussion of the manipulation of spin observables in accelerators or storage rings typically involves the Thomas-BMT [30] [31] equation in the absence of electric fields

$$\frac{d\vec{S}}{dt} = \frac{q\vec{S}}{m\gamma} \times [(1+G\gamma)\vec{B}_T + (1+G)\vec{B}_L] \quad (2.1)$$

This equation is consistent with Lorentz invariance and rotational invariance combined with helicity conservation and finite symmetries as discussed in the introduction provided we pay attention to the separation of transverse and longitudinal components of the Wigner Pauli-Lubanski tensor. To see this we consider the "natural" noninertial coordinate system

$$\vec{r} = \vec{r}_o + x\hat{e}_x + l\hat{e}_l + y\hat{e}_y \quad (2.2)$$

defined with respect to a reference orbit in a storage ring containing the point $\vec{r}_o$. At any point in the orbit, this noninertial reference frame is chosen such that it rotates with the local cyclotron frequency

$$\vec{\omega}_c = -\frac{q}{m\gamma}\vec{B}_T \quad (2.3)$$

In this rotating reference frame, we can write (2.1) in the form [31]

$$\frac{d\vec{S}}{dt} = (\vec{\omega}_s - \vec{\omega}_c) \times \vec{S} = -\frac{q}{m\gamma}[G\gamma\vec{B}_T + (1+G)\vec{B}_L] \times \vec{S} \quad (2.4)$$

In doing so, we have separated the transverse and longitudinal degrees of freedom for the particles in the beam so that the spin 3-vector $\vec{S}$ in this equation can now be identified with $\vec{S}_T = \frac{1}{m}\langle W^\mu \rangle_T$ as defined in eq. (1.15). The gauge invariance, boost invariance and rotational invariance of (2.4) can then be verified by a simple calculation. When a particle in the beam passes through a section of transverse magnetic field, its transverse spin vector rotates around the direction $\hat{n}$ by an angle

$$\delta\theta_T = \frac{qG\gamma}{mv\gamma}\int \vec{B}_T \cdot \hat{n}\, dl \qquad (2.5)$$

Similarly, when a particle passes through a section of longitudinal magnetic field, its transverse spin vector rotates around $\hat{e}_l$ by an angle

$$\delta\theta_L = \frac{q(1+G)\gamma}{mv\gamma}\int \frac{\vec{B}_L \cdot \hat{e}_l\, dl}{\gamma} \qquad (2.6)$$

The ratio of these two rotations is determined by the gauge invariance and Lorentz invariance of electromagnetism.

$$\frac{\delta\theta_T}{\delta\theta_L} = \frac{G}{1+G}\frac{\int \vec{B}\cdot\hat{n}\, dl}{\int \frac{1}{\gamma}\vec{B}\cdot\hat{e}_l\, dl} \qquad (2.7)$$

Gauge invariance requires that the Dirac moment rotate with the momentum so that the spin rotation in a transverse magnetic field is given by the anomalous moment while a rotation from a magnetic field parallel to the momentum engages the full magnetic moment. In addition, magnetic field strengths transform under boosts such that the boost that transforms the field strengths from the lab frame to the particle rest frame gives

$$\Gamma:(\vec{B}_T, B_L) \to (\vec{B}_T, \frac{1}{\gamma}B_L) \qquad (2.8)$$

Local rotational invariance requires that the magnitude of transverse rotations does not depend on the orientation of $\hat{n}$ in the plane normal to the momentum. This requirement is thoroughly tested, for example, by helical magnets that are used to make Siberian snakes or spin rotators. Therefore, the fact that eq (2.4) meets these constraints indicates that it represents a proper separation of the longitudinal and transverse dynamical degrees of freedom. It is also apparent that naïve application of the Thomas BMT equation without proper attention to the separation of longitudinal and transverse spin observables can lead to unintended parity violations or Lorentz noninvariant results. The thing to keep in mind is that the difference in the RHS of equations (2.1) and (2.4) describes the Wigner rotation between a fixed reference frame and the natural frame.

As mentioned earlier, we are interested in the case of a storage ring where the main bending magnets give magnetic fields in the $\hat{e}_y$ direction such that the center of the beam travels in the x-z plane. We now consider an additional magnet for spin manipulations. In a straight section of the ring, we locate an RF dipole magnet with the center of the magnet located at $\vec{r}_o$ as given in (2.2). When this magnet is turned off, natural reference frame reduces to Frenet-Serra frame and the local segment of the natural coordinates comprise an inertial frame:

$$B_x = 0 \rightarrow (\hat{e}_x, \hat{e}_l, \hat{e}_y) = (\hat{e}_x^I, \hat{e}_z^I, \hat{e}_y^I) \tag{2.9}$$

where the superscript I denotes inertial. When the magnet is turned off, we assume that the stable operation of the ring produces a beam of particles with a mean 4-momentum per particle in this local segment of the ring that can be given in this local inertial frame as

$$\langle P^\mu \rangle = (E, 0, 0, P_z) \tag{2.10}$$

This means that the vector map given by $\vec{\omega}_c$ in eq. (2.4) is quasi-periodic with a vanishing mean displacement. However, with the dipole magnet turned on we have a local magnetic field

$$\vec{B}_T = \hat{e}_x B_x \cos(\omega_m t + a) \tag{2.11}$$

where both the magnitude, $B_x$, and the RF frequency, $\omega$, of the magnetic field are controlled by the experiment. The operation of the magnet produces a coupled motion involving both the momenta of the particles in the beam and their spins. The local segment of the natural noninertial frame now oscillates with respect to the local inertial frame defined in (2.9). For small enough values of $B_x$ the vector map given by $\vec{\omega}_c$ remains quasiperiodic but the map now includes additional displacements in momentum. Neglecting the effect of other oscillatory motion within the beam, the operation of the RF magnet induces a momentum precession around the direction $\hat{e}_x$ that can be written in the local inertial frame in the form

$$\langle P^\mu \rangle_B \cong (E', 0, \delta p_y \sin[(\omega_m - \omega_c)t + a_y], P_z + \delta p_z \cos[(\omega_m - \omega_c)t + a_z]) \tag{2.12}$$

This approximate expression assumes that the perturbation of the beam momenta produced by the local magnet is small enough that no higher-order effects are generated in the remainder of the ring. The values $\delta p_y$ and $\delta p_z$ are then limited by the focusing elements of the ring. This can be true for small enough $B_x$ and for values of the RF frequency far from any resonance conditions of the ring. The validity of these stability requirements can be determined by measuring beam parameters during the operation of the dipole magnet at fixed frequency. If these parameters are stable, the dipole magnet can be defined as operating in a safe environment for spin manipulation. The variations shown in (2.12) will then average to zero. Although the mean direction of momentum will not change, the variation (2.12) will also result in forced betatron oscillations and forced synchrotron oscillations that will be addressed in Section IV. To calculate the "intrinsic" or "natural" component of the induced resonance strength we will neglect these effects.

As discussed in the introduction, the parity invariance of electrodynamics will ensure helicity conservation so that the longitudinal components of the Wigner-Pauli-Lubanski tensor will track with the momenta. The term in the Thomas-BMT equation (2.1) proportional to $\vec{\omega}_c \times \vec{S}$ that is associated with the helicity will thus lead to changes that also average to zero. In contrast, a Froissart-Stora frequency sweep [8] with the dipole magnet from $\omega_{initial} = \upsilon_s - \Delta f$ to $\omega_{final} = \upsilon_s + \Delta f$ will coherently rotate $\vec{S}_T = \frac{1}{m}\langle W^\mu \rangle_T$ through a finite angle. To see this cancellation of the oscillating terms explicitly, it is helpful to work in the lab frame and to parameterize the Wigner rotation, $\delta\varphi_W$, about $\hat{x}$ that defines the connection between the natural reference frame defined above and the lab frame using (2.12). Without loss of generality, we can consider a beam density matrix consisting of a single eigenstate of $\Sigma_y$,

$$\rho_\psi = |\psi\rangle\langle\psi| \qquad (2.13)$$

In the noninertial, natural frame, the momentum appropriately decouples from the transverse spin degrees of freedom and we can therefore write the 1-turn rotation matrix for $|\psi\rangle$ in a fixed reference frame in the form

$$|\psi\rangle_{n+1} = \exp\{2\pi i[\upsilon_s \hat{e}_y \cdot \vec{\Sigma} + (qG\gamma I_{Bx} \cos\omega(t_{n+1} - t_n) + \delta\varphi_W)\hat{e}_x \cdot \vec{\Sigma}]\}|\psi\rangle_n \qquad (2.14)$$

where $I_{Bx} = \int B_x dl$ and the $\vec{\Sigma}$ are in a representation of the SU(2) or SO(3) algebra given in (1.20) that is appropriate for the type of particle beam being rotated. For a discussion of the vector polarization, it is sufficient to consider a spinor representation and the equation (2.14) for the vector polarization can then be written in the form

$$\frac{\partial|\psi\rangle}{\partial s} = -\frac{i}{2}\begin{pmatrix} \nu_s & r(s) \\ r^*(s) & -\nu_s \end{pmatrix} \qquad (2.15)$$

Where we can use (2.12) to write

$$r(s) = -2i\varepsilon_K^D\left[\cos(\nu_m s) + (\delta p_y/G\gamma)\sin((\nu_m - 1)s)\right] \qquad (2.16)$$

To extract the residue, it is convenient to transform $|\psi_R\rangle = e^{-i\nu_m s\sigma_3/2}|\psi\rangle$ in order to write eq. (2.15) in the form

$$\frac{\partial|\psi_R\rangle}{\partial s} = -\frac{i}{2}\begin{pmatrix} \nu_s - \nu_m & r'(s) \\ r'^*(s) & \nu_m - \nu_s \end{pmatrix} \qquad (2.17)$$

with the connection

$$r'(s) = -i\varepsilon_K^D\left(1 + e^{2i\nu_m s}\right) + \varepsilon_K^D(\delta p_y/G\gamma)\left(e^{is} - e^{i(2\nu_m - 1)s}\right) \qquad (2.18)$$

The oscillating terms in this expression average to zero with two different time scales. The fact that the second term, proportional to $\delta p_y$, averages to zero is guaranteed by the beam stability assumptions mentioned above. There can be no net deflection of momentum in the vertical direction. In a region where the oscillating terms in the off-diagonal elements average to zero, this then gives

$$\frac{\partial |\psi_R\rangle}{\partial s} = -\frac{i}{2}\begin{pmatrix} v_s - v_m & -i\varepsilon_K^D \\ +i\varepsilon_K^D & v_m - v_s \end{pmatrix}|\psi_R\rangle \tag{2.19}$$

A simple comparison of (2.5), (2.14) and (2.19) then gives the resonance strength

$$\varepsilon_K^D = \frac{qG\gamma}{2\sqrt{2}\pi p}\int B_T^{rms} dl \tag{2.20}$$

for a controlled Froissart-Stora spin rotation where $v_m$ sweeps adiabatically through $v_s$. Bai, MacKay and Roser, in Ref. [32] demonstrated clearly that improperly including the oscillating off-diagonal elements proportional to $e^{2iv_m s}$ can give an incorrect additional factor of two in the expression for the resonance strength. Unfortunately, they did not consider spin rotations in the "natural" reference frame that separates longitudinal and transverse degrees of freedom and did not recognize that the term proportional to $1/G\gamma$ represents a Wigner rotation that also averages to zero. The thus achieved a result that corresponds to the conventional expression given in (1.27). This separation is not unimportant. Exactly the same arguments that, when applied to a ring with only bending magnets with fields in the $\hat{e}_y$ direction, give the spin tune

$$\upsilon_s = G\gamma \tag{2.21}$$

and <u>not</u> $(1+G\gamma)$ or $(1+G)\gamma$ combined with local rotational invariance are involved in the derivation of (2.20). Note that it is extremely important when considering the local transverse spin advance to treat the spin and momentum deflections in a consistent manner.

References [29] and [32] present derivations of the intrinsic or natural resonance width associated with the induced resonance produced by a localized RF dipole field in a controlled Froissart-Stora sweep. However, these papers do not consider the separation of longitudinal and transverse spin observables and, consequently, their calculations do not respect gauge invariance or parity conservation. Each paper gives an expression for the local rotation of "spin" that treats rotations around $\hat{e}_y$ as being different from rotations around $\hat{e}_x$. In addition, these papers do not define spin orientation in terms of a measurable property of a polarized beam related to the Wigner-Pauli-Lubanski tensor and their derivations of the natural resonance width for a local dipole field coincide with the "conventional" expression given in Eq. (1.27). The difference between (1.27) and the Kondratenko natural resonance strength as given by (1.29) or (2.20) vanishes as $\left(\frac{1}{\gamma}\right) \to 0$ but can be important, particularly for the spin manipulation involving electrons and deuterons. In spite of their omissions, references [29] and [32] both make the very important point that the intrinsic or natural resonance strength is not the only contribution to spin rotation in a Froissart-Stora sweep with an RF dipole but must be considered in conjunction with other contributions. To illustrate this point, we now turn to a discussion of the impact of momentum oscillations and the "forced" component of spin rotations.

### III. Natural Resonance Strength and Forced Resonance Strength

The derivation presented above is valid for the natural resonance strength induced by a localized dipole providing the strength of the magnet is small enough so that certain consistency conditions (2.12) for local momentum measurements on the beam are met. This derivation is not valid for a ring containing Siberian snakes since the one-turn spin rotation matrix in those situations is more complicated. The expression for the resonance strength given by (1.29) or (2.20) depends only on the local RF magnetic field and can properly be termed an intrinsic or natural resonance strength. In addition, as mentioned above, it is not the only contribution to the resonance strength for a coherent spin manipulation involving a Froissart-Stora sweep.

As shown by Bai, MacKay and Roser [32] and independently by S.Y. Lee [29] when considering spin manipulations by RF dipole fields, it is also important to consider the impact produced by the RF magnet on forced betatron motion in the ring. Lee denotes the component of resonance strength associated with this effect the "forced resonance strength". By definition, the forced resonance strength depends on other ring parameters and its calculation therefore depends on these parameters. We will follow here the approach of Lee [29] for understanding how the component of resonance strength associated with forced betatron oscillations occurs.

In the presence of an operating RF dipole magnet, the equation for vertical deflection from a flat trajectory in an accelerator can be written

$$\frac{d^2 y}{dl^2} + \kappa(l) y = I_x \cos(\omega l / v + \chi) \sum_n \delta(l - nC) \qquad (3.1)$$

where $I_x = \int B_x dl$, $\kappa(l)$ is focusing function and C is the effective circumference of the ring. The Floquet transformation involves the substitutions

$$\eta = y/\sqrt{\beta}, s = \frac{1}{\nu_y}\int_0^l \frac{dl}{\beta} \tag{3.2}$$

and leads to the equation

$$\frac{d^2\eta}{ds^2} + \nu_y^2 = \frac{\nu_y \beta_o^{\frac{1}{2}} I_x}{2\pi} \sum_n \cos[(n+\nu_m)s + \chi] \tag{3.3}$$

in which $\beta_o$ is the betatron oscillation amplitude at the location of the RF dipole magnet. The solution to this inhomogeneous differential equation can be written in the form

$$\eta = \eta_o + a_1 \cos(\nu_y s) + a_2 \sin(\nu_y s) + \eta_f \tag{3.4}$$

where the forced oscillation term is given by

$$\eta_f = \sum_n \frac{\nu_y \beta_o^{\frac{1}{2}} I_x}{2\pi \left[\nu_y^2 - (n+\nu_m)^2\right]} \cos\left[(n+\nu_m)s + \chi\right] \tag{3.5}$$

For the treatment of spin rotation involving a controlled Froissart Stora sweep with $\nu_m \to \nu_s$ we can see that the forced betatron oscillations have the effect of enhancing the 1st –order intrinsic spin resonances located at

$$\nu_s^{n\pm} = n \pm \nu_y \tag{3.6}$$

Since the strength of intrinsic spin resonances depends on the magnitude of oscillations, this enhancement depends on the magnitude and the location of the dipole magnetic field $I_x = \int B_x dl$ and on other accelerator parameters. For the rotation from a frequency-controlled Froissart-Stora sweep that <u>only</u> crosses the location of the induced resonance $\nu_m = \nu_s$, the enhancement of the numerous intrinsic spin resonances thus has the effect of modifying the effective resonance strength for the induced spin resonance

$$\varepsilon_{FS}^D(\nu_y) = \left|\varepsilon_K^D(1 + g_B(\nu_y))\right| \tag{3.7}$$

From (3.5) we see that the functional form of $g_B(\nu_y)$ depends both on the local magnetic field and on the specific details of the accelerator lattice in the experiment. At a given value of $\nu_y$, this modification can be either positive or negative. The absolute value appears in this equation because the resonance strength extracted from a Froissart Stora sweep (1.25) is always chosen to be the positive square root of a positive number. In the approximations leading to (3.5) it can be seen that $g_B(\nu_y)$ is a meromorphic function of $\nu_y$ and that its Mittag-Leffler expansion can be written in the form

$$g_B(\nu_y) = \sum_n \left\{ \frac{c_n^-(B_x, \nu_y)}{\nu_y - \nu_s + n} + \frac{c_n^+(B_x, \nu_y)}{\nu_y + \nu_s + n} \right\} \tag{3.8}$$

S.Y. Lee, in Ref. [29], describes a detailed program for calculating the residues of the poles in this expression. However, the discussion in that paper makes the assumption that the natural resonance strength is given by the "conventional" expression (1.27). The Lee approach to the challenge of calculating forced resonance strengths has also been extensively explored by A. Lehrach [33] using the known parameters of the COSY ring. The detailed comparison of these studies with specific measurements can be very interesting and can provide valuable information about beam dynamics. However, in performing these studies, it is obviously important to have a precise definition of the intrinsic resonance strength of the RF magnet doing the perturbations. This is one of the reasons for the calculation leading to (2.8) presented here.

The value of considering, in detail, the full range of dynamical effects generated by a localized RF magnetic dipole field has been appreciated by others. For example, Kondratenko [34] has recently addressed these forced betatron and synchrotron oscillation effects in some detail using sophisticated analytic techniques. In addition, Barber [35] has advocated a comprehensive approach focused on the application of explicit computer models for beam dynamics. Our brief presentation here has concentrated only on the impact of forced betatron oscillations in order to illustrate some of the issues involved. Although the synchrotron oscillations can also have a finite effect on coherent spin manipulations, we will not address separately their collective impact on the effective resonance strength, (3.7), for a Froissart-Stora sweep. We now turn to a brief discussion of some of the implications of experimental results involving coherent rotations as a function of betatron tune.

### IV. Effective Froissart-Stora Resonance Strengths for RF Dipoles as a Function of Betatron Tune

The subject of coherent spin manipulations using localized RF magnetic fields has considerable practical experimental interest. The most direct demonstration of the application of these techniques involves the successful efforts of the Brookhaven accelerator physics team (described by Bai et al. [23]) to employ an RF dipole in order to enhance the strength of intrinsic spin resonances in the AGS so that the combination of the two coupled resonances could be crossed with a spin flip and minimal loss of polarization for subsequent injection into RHIC. In other venues, the "Phelps" technique [9] of using the induced resonance produced by an RF magnetic field to coherently rotate a polarized beam of fixed energy was shown to provide varied opportunities to reduce systematic experimental errors in the measurement of spin observables in experiments in storage rings. Early studies [9-17] of these controlled Froissart-Stora sweeps concentrated on achieving maximal spin flip efficiency. However, the compilation by Leonova et al. [36] of the experimentally-determined effective resonance strengths involving both RF solenoid and RF dipole sweeps directed attention to important quantitative aspects of coherent spin rotations.

One of the original aims stated for the Leonova compilation was to address a factor of two discrepancy in the available values [27] [37] for the natural resonance strengths associated with both solenoid and dipole RF magnetic sweeps. The widely scattered set of values uncovered by the Leonova compilation was not able to provide useful information about this issue but the work of Bai, MacKay and Roser [32] and the work of Lee [29] conclusively settled the erstwhile "factor of 2" dialogue and correctly directed attention on the difference between solenoid and dipole resonance strengths associated with the additional forced betatron and synchrotron oscillations produced by RF dipole fields.

It is important to observe that the separation of the effective resonance strength for an RF dipole sweep into its natural or intrinsic component and its forced component as defined in Ref. [29] can be studied quantitatively by doing controlled Froissart-Stora sweeps for different values of betatron tune. In this context, it is instructive to consider (3.7) with the betatron tune in the vicinity of an intrinsic resonance. This can be written in the form,

$$\varepsilon_{FS}^D(\nu_y) = \left| \varepsilon_K^D \left(1 + \tilde{g}_b(\nu_y)\right) + \frac{\varepsilon_K^D c_R(B_x, \nu_y)}{\nu_y - \nu_R} \right| \quad (4.1)$$

In this equation, the function

$$\tilde{g}_B(\nu_y) = g_B(\nu_y) - \frac{c_R(B_x, \nu_y)}{\nu_y - \nu_R} \quad (4.2)$$

is obtained by removing the specific resonance being studied from the Mittag-Leffler expansion, (3.8). Note that there is necessarily a value of the betatron tune

$$\nu_y^o = \nu_R - \frac{c_R(B_x, \nu_y)}{1 + \tilde{g}_B(\nu_y)} \quad (4.3)$$

for which the effective resonance strength for a sweep involving only the RF induced resonance vanishes. This shift provides the value of the residue associated with this pole in terms of its contribution to the expansion (3.8). The vanishing of the effective resonance strength is a result of the coupling of the induced resonance and the intrinsic resonance as described in Ref. [23]. The vanishing of the effective resonance strength also provides an explicit example of a result required by a rigorous theorem for resonance crossing proved in Ref. [26]. Because of the absolute value in (4.1) and the spread in betatron tune for a given polarized beam setup, the search for $\nu_y^o$ reduces to looking for the minimum of the measured resonance strength. Note that the location of this minimum determines the sign and the magnitude for the residue of this specific pole. Since the strength of an in intrinsic resonance is proportional to the betatron oscillation amplitude, eq. (3.5) shows that it is useful to study interference (4.1) as a function of the magnitude of dipole magnetic field. In a controlled Froissart-Stora sweep, this study can be done by adjusting $B_x$ and the ramp parameters so that

$$\left(\varepsilon_K^D\right)^2 \frac{\Delta t}{\Delta f} = const. \tag{4.4}$$

in eq. (1.25). As this is done, we should have a region where the forced oscillations reduce in proportion to the natural betatron oscillations so that $\tilde{g}(v_y) \to 0$ and we have

$$\varepsilon_{FS}^D = \varepsilon_K^D \left(1 + \frac{c_R}{v_y - v_R}\right) \tag{4.5}$$

where $c_R$ is the normal strength of the intrinsic resonance.

This simple example has been discussed in order to illustrate the manner in which coherent spin manipulations can be used to explore details of beam dynamics. The more comprehensive treatments using the techniques developed by Lee [29], Kondratenko [34] provide a wide variety of other examples. The detailed computer studies advocated by Barber [35] may uncover unexpected gems. However, the utility of these quantitative studies depends on a correct treatment of local gauge invariance. Spin is a unique property of massive particles and spin observables for polarized beams reflect the well-known properties of the Wigner-Pauli-Lubanski tensor for specific particle types. The Kondratenko conjecture [28] for the resonance strength induced by the operation of an RF dipole magnet demonstrates that these constraints must be respected in order to avoid unnecessary errors. The excellent summary "First Polarized Proton Collisions at RHIC" by T. Roser et al., [38] gives a fine introduction to local rotations of transverse spin observables that illustrates the issues discussed here in the context of helical spin rotators and Siberian snakes.

Two studies of the behavior of the effective resonance strength using an RF dipole magnet as a function of betatron tune have been published by the Spin @ COSY collaboration. The experimental papers, [36,39] should be consulted for results. The discussion in this section is intended to be an aid in shaping future studies.

**Acknowledgements**